# Physical Access Control Management System Based on Permissioned Blockchain

Sara Rouhani
Department of Computer Science
University of Saskatchewan
Saskatoon, Canada
sara.rouhani@usask.ca

Vahid Pourheidari
Department of Computer Science
University of Saskatchewan
Saskatoon, Canada
vahid.p@usask.ca

Ralph deters
Department of Computer Science
University of Saskatchewan
Saskatoon, Canada
deters@cs.usask.ca

*Abstract*— Using blockchain as a decentralized backend infrastructure has grabbed the attention of many startups entrepreneurs and developers. Blockchain records transactions permanently and protects them from undesirable tampering. It provides a reliable tamper-proof database which can be considered as a trustable source for tracking the previous system state. In this paper, we present our access control application based on Hyperledger Fabric Blockchain and Hyperledger Composer to control access to physical places. The system components and modular architecture are illustrated, and we have extracted metadata include historian transactions' details arising from our demo test. Finally, the performance metrics and resources consumption are provided using Hyperledger Caliper, a benchmark framework for measuring Hyperledger blockchains performance.

## I. INTRODUCTION

Access Control systems are applied in computer security to specify the conditions and implementation of granting or revoking access to logical or physical resources. Applying blockchain for access control systems is one of the in-demanding areas for using this new technology. Blockchain's unique characteristics such as durability, immutability, reliability, auditability, and security, made it a perfect solution for access control systems for a variety of applications such as healthcare systems [1]. Blockchain can be considered as a distributed database to manage access control and includes the permanent history of transactions which are tamper-proof and indelible. In addition, blockchain removes the third parties' part from the system and consequently it provides privacy.

There are two types of access control systems, Logical Access Control, and Physical Access Control. Most of the studies on access control systems focus on logical access control, however, we have considered blockchain's potential as an infrastructure for access control management software on the top of the physical access control systems. Blockchain provides a tamper-proof solution for access control systems which provides an auditable trail of transactions. There are two advantages of using blockchain for physical access control systems. First the history of all recorded transactions stores on a reliable tamper-proof database, hens significantly reduces the chance of collusion. Second management of access rights performs by a distributed approach, so the role of the third party eliminates from the system and authorized users can grant or revoke another users' access directly.

In this research study, we have implemented an access control application based on Hyperledger Fabric permissioned blockchain. We have implemented a simulated scenario to test the application, and the metadata has been extracted from the historian record of the submitted transactions.

In the following of the paper in section two, we overview the background related topics. We review the related works in section three. Section four presents our system components and architecture model. Analysis of our application available in section five and finally, section six states summary and future works.

## II. BACKGROUND

### A. Blockchain and Smart Contracts

Blockchain initially introduced by bitcoin [2], a successful decentralized system for transferring digital currencies directly. Blockchain is the new generation of the distributed database systems and it is handled by the peer to peer network. It is the chain of chronological blocks of transactions which is submitted by untrusted parties and they confirm without any central authority. The chain starts with genesis block and it continuously grows when a new block links to the previous block through a hash value of the previous block. In other words, for hash value calculation of any new block, the hash value of the previous block is considered as well. That feature makes blockchain tamper resistant because if someone attempts to make a change, the hashes for linked blocks will also change and this disrupts the ledger state and this makes the conflict between the copies of the shared ledger. The concept of shared ledger refers to this point that every peer on the blockchain would have access to a copy of the whole ledger. After changing to the blockchain unique state, the process of syncing starts and again every peer will access the updated state of the blockchain.

Auditability is one of the key characteristics of the blockchain. After that each transaction validated, it records on the current block with a timestamp and blockchain's users are able to track the previous transactions and they can access the history of all transactions. This feature is very interesting from data management perspective and the applications that are required to access to tamper-proof log history.

Blockchain uses different consensus algorithms to reach agreement on the new state for the blockchain. The main idea of using consensus algorithm is to use a group of people who are involved in the system for decision making instead of using a trusted third party for making decisions. Proof of work

(PoW), proof of stake (PoS) and Practical Byzantine Fault Tolerance (PBFT) are the most known cases of blockchain consensus algorithms. These consensus algorithms are different in identity management mechanism, energy saving, and tolerating power of adversary [3].

Smart Contract is a set of programmable functions that store on the blockchain and automatically enforce its terms without the need for trusted intermediaries. Smart contract's functionality is almost like traditional contracts. As the traditional contracts include conditions that each of the parties is obligated to follow, blockchain smart contracts have been designed to automate the same task by automatically validating conditions and running the next steps automatically based on the result of the conditions. However, smart contracts automate transactions without the need for a central authority or legal system. By utilizing smart contracts, we can define more complex transaction tangled under specific conditions on the blockchain and as a result, we can develop application beyond just transferring digital currencies in more various are such as supply chain, business process management, and healthcare.

## B. Public and Private Blockchain

Bitcoin is the first creation of blockchain and it is public. That means everybody with anonymous identity can join to the blockchain, read the blockchain, send transactions and be considered as a member of the consensus committee. Public blockchains may seem interesting from some application perspective since they are open to the world and their users' identity is not known, however, the emerging of private blockchains are tailored from intra organization perspective to rolling out blockchain into diverse productions.

In permissioned or private blockchain, there is an additional permission layer to authenticate users who want to join to the blockchain. So, the main difference between public and private blockchain is who can be part of the system.

In addition, it is worth to mention that there is a third type of blockchain known as consortium blockchain and it could be considered as a hybrid type as only a selected of nodes can participate in decision making, and the access to read or wright on the blockchain could be public or restricted. [4].

Hyperledger is a set of projects for open-source industrial blockchain frameworks such as Fabric, Sawtooth, Burrow, and Iroha hosted by Linux Foundation.

Hyperledger Fabric provides modular architecture and includes Membership component to create flexible permissioned blockchain platforms. Fabric supports smart contracts called "chaincode" which is programming code for implementing the application logic and transaction functions. Fabric overcomes the problems of permissioned blockchain such as non-deterministic execution of concurrent transactions, execution on all nodes, non-flexible trust model, hard-coded consensus and etc. by using execute-order-validate architecture instead of order-execute architecture [5].

Fabric's ledger includes two components: World State and Transaction Log. The World State addresses the current state of the ledger and the Transaction log includes the history of all transactions. If a transaction changes in any value already stored on the ledger or adding new data to the ledger, this is considered as new state for the blockchain and it would be kept permanently, and it is not possible to reverse the previous state of the blockchain [6].

## C. Access Control Systems and blockchain

Access control refers to any action that restricts access to a physical place or logical resource. Physical Access Control systems have three main components: access control, surveillance, and testing. Physical access control investigates solutions to regulate access to physical places and physical resources for companies that want to manage access control over their private and sensitive physical resources. Such as using door lock restriction to access the important rooms. Logical access control refers to managing access control for software systems and data.

Current access control systems often focus on logical access control and they store data on a centralized server, which would be handled by server administrators, who have full access to everything and the access control processes are executed centralized. As a result, tampering always threat data and history of transactions. Even for many systems, it is not possible to detect data modification or data removal from the system.

Using Physical access control implementation across a variety of organization is getting more popular every day and multiple companies are developing cutting-edge technologies to handle secure physical access control technology. For example, a physical Access Control system based on the smart card is composed of three elements: smart card, door reader to check the card validation and the door or gate, which would be unlocked when the card is approved. However, behind the scenes of this hardware equipment, there is a complex network of data, servers, and software to authorize users and manage access control [11].

We found blockchain as an improvement for storing and managing software layer of physical access control. Using blockchain enables us to manage access control distributed and be able to access to the trusted history logs permanently. This trust comes from the strong and secure logic behind of the blockchain architecture instead of non-logical trusting centralized system administrators. Store permission and denial access transactions on blockchain provided reliable transaction logs to refer at any point in the future. In this paper, we introduce our noble blockchain based system to implement physical access control. Figure 1 shows how physical access control software management interacts with blockchain to request data from blockchain and read access permissions from blockchain.

For access control method we have used both role-based access control and rule-based access control method. We have used Role-Based access control method to regulate users access based on their role in the system. Also, there are some rules in the system that defines who has the privilege to change other users access permissions by submitting transactions. All of the transactions remain in blockchain database permanently, and nobody can delete a record of it. For implementing role-based access control we have used ACL module of

Hyperledger composer, by defining participants default access permissions. Each participant represents a different role in the system.

For dynamic access control, an authorized user who accesses to submit grantAccessControl or revokeAccessControl transactions submit appropriate transaction and changes related user access permissions when the transaction has been confirmed.

## III. RELATED WORKS

Medrec [1] was one of the initial studies which implement an access control system for sharing medical data based on Ethereum Blockchain. The system gives the authority of sharing medical data to the patients as the real owners of data. Medrec uses mining reward solutions to motivate medical stakeholders to participate in the system and verify transactions as miners. MeDShare [7] is another blockchain-based framework for sharing big medical data among cloud services to achieve data provenance and auditing.

[8] represents a blockchain based access control application which uses a bitcoin public blockchain to express the rights exchanges. Since it uses public blockchain everybody can see who has access to which resource. It provides auditability and security while it is compromising privacy.

[9] are blockchain-based access control studies which consider access control in IoT context. Their decentralized solution addresses the security and privacy issue related to the data provided by IoT devices.

## IV. SYSTEM IMPLEMENTATION

We used Hyperledger Composer framework [10] to implement our system on the top of the Hyperledger Fabric.

All components can be defined by the composer modular structure and then it can be packed as one component and deploy on the fabric blockchain. The main parts of the composer module are set of model file, an Access Control Language file, Query file and a set of JavaScript files.

In the model module, we have defined the participants, Assets, Transactions, and Events.

All users who use the system such as administrator users who define the access control policies and other authenticated users who need access to physical places have been defined as a participant in the model file. Defining participants include the participant unique key and its attributes.

We consider physical places as an asset in our system which access to them will be managed through transaction processing functions or smart contracts and access control file. For example, in our application, we consider a unique string ID which refers to the input door of physical places which secured by physical access control devices. It could include any other related attribute or map to any specific participant which can manage access policies over it.

Transactions are submitted to change the state of the blockchain based on their predefined conditions. The attributes of the transactions define in the model module, and their processing functions define in the JavaScript file. In our application, we have a transaction for granting access permissions and we have another transaction to revoke access from users. In addition, there is a transaction to delegate authority to other users, then they would be able to submit the grant access transaction to physical places to other users.

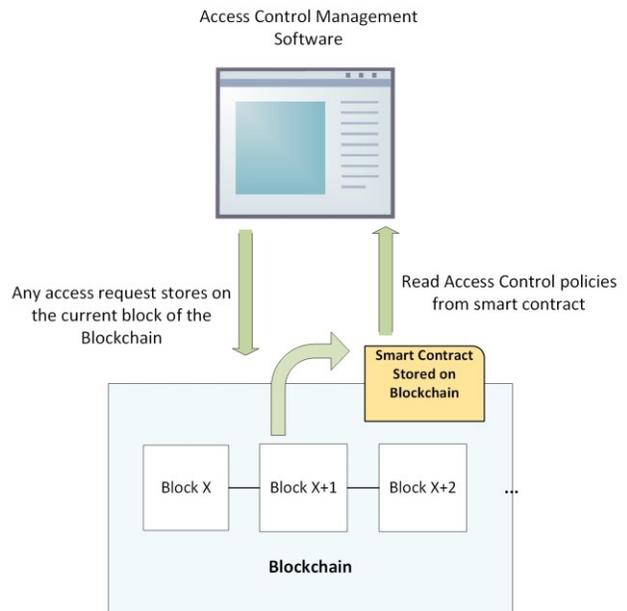

Figure 1. User Interface interaction with blockchain

We have defined our access control policies in the Access Control Language (ACL) module. Here is the list of rules we consider for our application. These policies include the following items.

- Participants with specific roles have access to which resources by default
- Participants with specific roles can send transactions
- System access permissions
- System administrator role access permissions
- Participants with specific roles access permissions to read historian records

Composer ACL language defines by five different sets, a set of Participants, a set of Resources, a set of Conditions, a set of Actions, and a set of Operations. Table one explains the meaning of each notation. In the basic model, users can access to the resources based on their role defined in the system, so our access control model is based on role-based access control model [13].

In our Composer model file, we have defined different type of participants which reflects different roles in the organization and they have access to resources by default based on their roles and conditions.

$Ur \subseteq U \times r$ is the set of user-role assignments and $rR \subseteq r \times R$ is the set of role-resource assignments.

TABLE I. NOTATIONS MEANING

| Notation | Meaning |
|---|---|
| P | Set of Participant |
| R | Set of Resources |
| C | Set of Conditions |
| A | Set of Actions (Create, Read, Update and Delete) |
| O | Set of Operations (Allow and Deny) |

A user u have access to the resource r if and only if there is a participant p defined in the model file and $(u,p) \in Ur$ and $(p,r) \in rR$ and condition c is met.

For a situation that an authorized user wants to change the access of any user, he or she can submit a transaction to change user access control dynamically and this is separated from static ACL module definitions.

Events are the important part of the system when they use alongside system queries. we have implemented event module in order to query the transactions information logs. Log entries demonstrate the results of the events have been fired from transactional processing functions. In addition, they can subscribe to an external application. For example, in our application, we consider situation insisting on access request and rejecting request after several consecutive efforts. As a result, a trigger to the external application will be fired to apply necessary safety considerations such as triggering intrusion alarms to deter unauthorized access.

Composer Transaction Processor functions are the part of JavaScript files and they translate to fabric chain code, so it can be considered as smart contracts. Their main task is to define the logic of each transaction and the conditions that need to meet. The Transaction Processor Functions are automatically called when the respective transaction is submitted. Figure 2 explains the access permission procedure based on ACL module

Composer Identity Management main tasks include create a participant and issue an identity for that participant. Composer uses the concept of Card inspiring from real-world ID cards to issue an identity to a participant. An ID card is access card to the chain and it includes identity data, a connection profile, and the certificate for chain access. Connection profile is used to connect composer network to the runtime Fabric. These ID cards can map to real-world physical access control smart cards.

Query file includes the declaration of queries. Queries based on tamper-proof data. We can be sure we query the data which has not been modified, altered or deleted.

Hyperledger Composer provides Historian record, which records successful transactions with details includes transaction's information and participant who submitted the transaction. Figure 3 illustrates the historian transaction details.

Figure 4 represents the architecture model of our application based on Hyperledger Composer and Hyperledger Fabric. It includes Model, JavaScript, Access Control, and query. All of these Modules merge into a Business Network Archive file and then the archive file (.bna) deploys on runtime Hyperledger Fabric Blockchain and afterward users can interact with the blockchain through JavaScript API and user interface application.

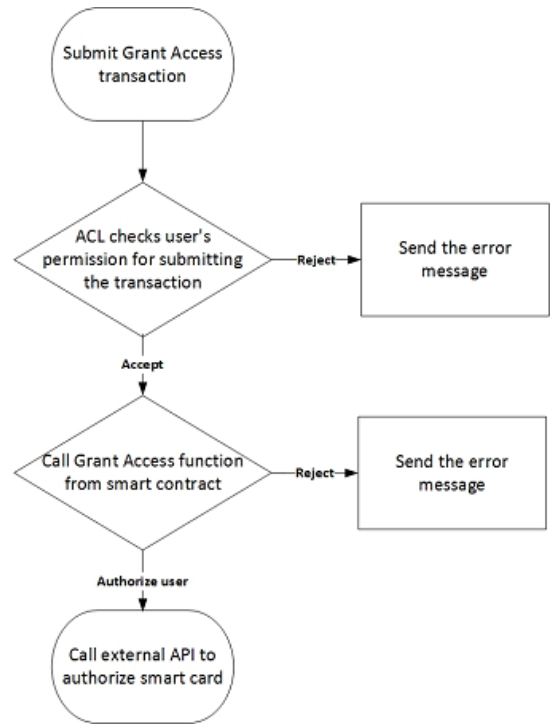

Figure 2. Access permission procedure based on ACL module

Although the performance is one of the main obstacles of blockchain-based applications, advancing in blockchain platforms by considering several factors promises a bright future for this issue [13]. In this paper, we have used Hyperledger Caliper [14] to test our application and measure performance metrics. Hyperledger Caliper is a framework to test different Hyperledger blockchains such as Fabric, Sawtooth, Iroha, and Composer and get a report includes set of performance indicators.

A test module for our application has been created to generate and submit transactions. Three main functions of our test module comprise init(), run(), and end() based on Caliper requirement. The init function handle the initialization phase of the test. The run function generates and submits multiple transactions repeatedly in an asynchronous way. Finally, the end function finalizes the end of the test.

V. APPLICATION ANALYSIS

In our model file, we have defined different participants which map to different roles in the system and they have a different level of access such as managers and employees. We considered asset as the physical places and we categorized them into different departments, so access to each department would be control by a different user (the CEO of the department). User or participant instances and asset instances have defined in init function. Afterward, test Transactions have been fired through run functions. The test result indicates that our application has worked 100% correctly during the test phase. Tables I and II present performance metrics and resource consumption reported by Hyperledger Caliper respectively.

## VI. SUMMARY AND FUTUREWORKS

In centralized systems access to resources control by third parties such as system administrators who have full control on system's data and actions, and as a result they always suffer from security and trust issues. This paper is the implementation of a real-world application using blockchain technology to explore permissioned blockchain applications. By exploiting Hyperledger Fabric and Hyperledger Composer potential we have implemented a tamper-proof access control application based on permissioned blockchain for managing access permissions on physical places.

The system provides a comprehensive transactions' log to query and it could be accessible through authenticated and authorized users. In addition, the transaction history of the system is trustable since it is protected from undesirable tampering. The analysis result reported by Hyperledger Caliper illustrates system stability and scalability include 100% successful transaction rate along with acceptable performance metrics.

As a future work, we intend to continue our research on access control management using blockchain technology to integrate logical and physical access control and use formal methods to analyze the security aspect and eliminate exploitable bugs.

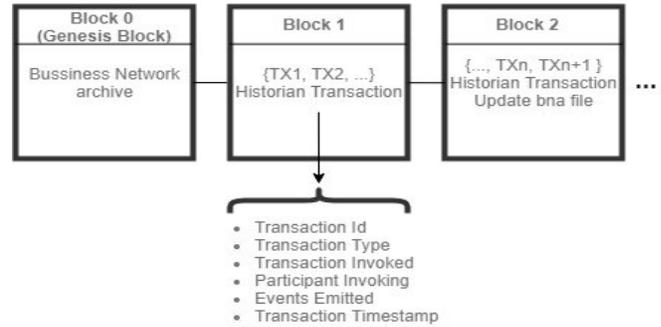

Figure 3. Historian Transaction Record

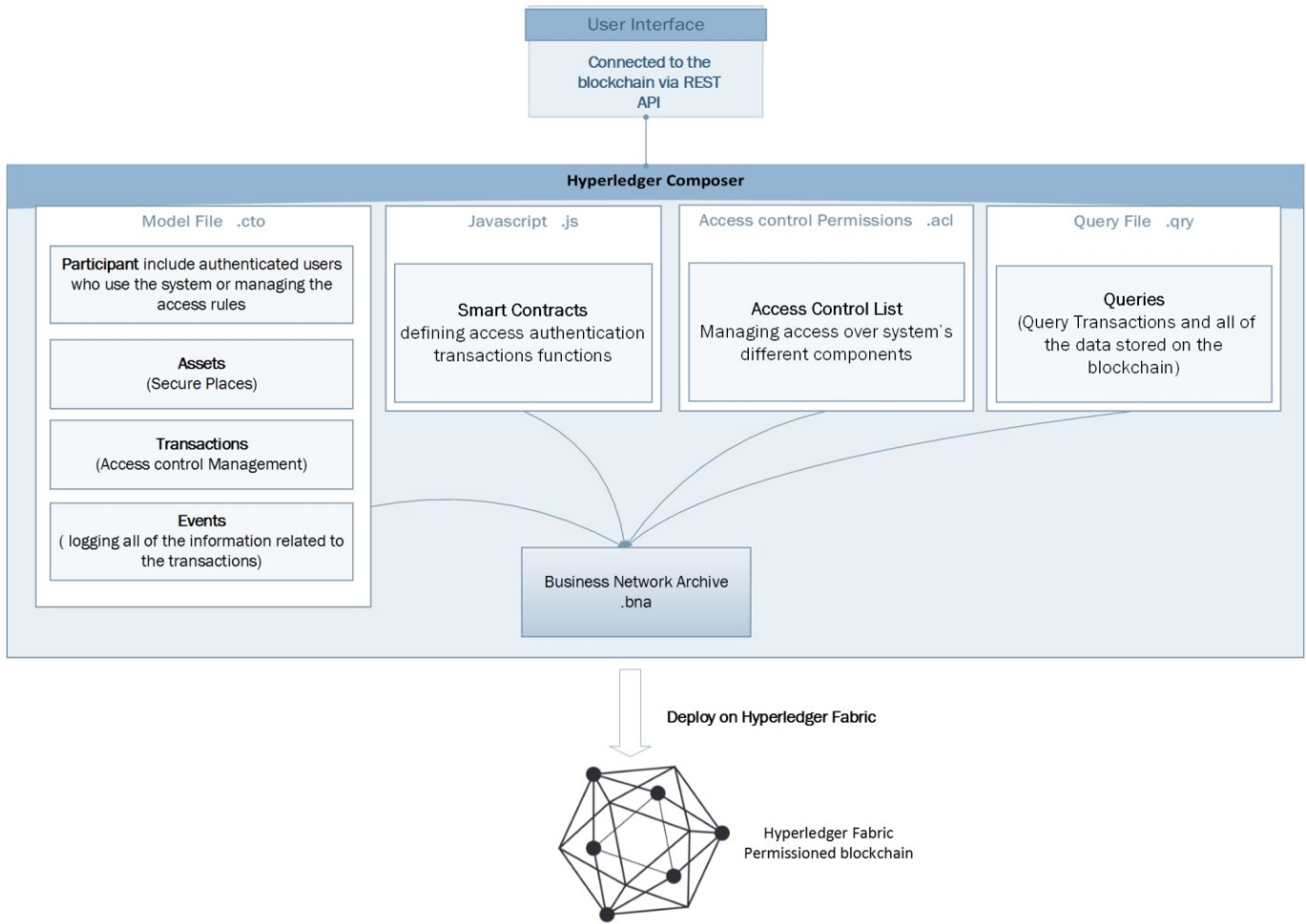

Figure 4. System Architecture

TABLE II. PERFORMANCE METRICS

| Test | Name | Succ | Fail | Send Rate | Max Latency | Min Latency | Avg Latency | 75%ile Latency | Throughput |
|---|---|---|---|---|---|---|---|---|---|
| 1 | physical-access-network | 500 | 0 | 10 tps | 20.67 s | 1.32 s | 13.72 s | 18.18 s | 7 tps |

TABLE III. RESOURCE CONSUMPTION

| TYPE | NAME | Memory(max) | Memory(avg) | CPU(max) | CPU(avg) | Traffic In | Traffic Out |
|---|---|---|---|---|---|---|---|
| Process | node bench-client.js(avg) | - | - | NaN% | NaN% | - | - |
| Docker | dev-peer0.org1.example.co...0.1.0 | 125.5MB | 121.7MB | 101.01% | 20.61% | 4.5MB | 3.7MB |
| Docker | dev-peer0.org2.example.co...0.1.0 | 121.4MB | 115.1MB | 102.06% | 21.12% | 4.5MB | 3.6MB |
| Docker | peer0.org1.example.com | 368.4MB | 343.5MB | 17.73% | 11.75% | 16.6MB | 32.5MB |
| Docker | peer0.org2.example.com | 359.6MB | 333.3MB | 17.87% | 11.66% | 16.5MB | 32.7MB |
| Docker | couchdb.org1.example.com | 114.5MB | 109.2MB | 54.88% | 32.39% | 4.5MB | 8.1MB |
| Docker | couchdb.org2.example.com | 117.2MB | 110.9MB | 53.61% | 32.86% | 4.5MB | 8.1MB |
| Docker | orderer.example.com | 17.7MB | 14.7MB | 3.64% | 1.85% | 3.9MB | 7.8MB |
| Docker | ca.org1.example.com | 5.3MB | 5.3MB | 4.37% | 0.20% | 4.6KB | 3.6KB |
| Docker | ca.org2.example.com | 7.3MB | 7.3MB | 0.00% | 0.00% | 1.9KB | 0B |